\documentclass{article}
\usepackage{arxiv}

\usepackage[utf8]{inputenc} 
\usepackage[T1]{fontenc}    
\usepackage{hyperref}       
\usepackage{url}            
\usepackage{booktabs}       
\usepackage{amsfonts}       
\usepackage{nicefrac}       
\usepackage{microtype}      
\usepackage{lipsum}		
\usepackage{subfigure}
\usepackage{graphicx}
\usepackage{indentfirst}
\usepackage{color}
\usepackage{algorithm}
\usepackage{amsthm}
\usepackage{algpseudocode}
\usepackage{multirow}
\usepackage{authblk}
\usepackage{doi}
\title{Deep Co-supervision and Attention Fusion Strategy for Automatic COVID-19 Lung Infection Segmentation on CT Images}

\date{}

\author[1,2]{Haigen Hu}
\author[1,2]{Qiu Guan}
\author[1,2]{Xiaoxin Li}
\author[1,2,*]{Qianwei Zhou}
\author[3,*]{Su Ruan}
\affil[1]{College of Computer Science and Technology, Zhejiang University of Technology}
\affil[2]{Key Laboratory of Visual Media Intelligent Processing Technology of Zhejiang Province}
\affil[3]{University of Rouen Normandy}

\hypersetup{
pdftitle={Deep Co-supervision and Attention Fusion Strategy for Automatic COVID-19 Lung Infection Segmentation on CT Images},
pdfauthor={Haigen Hu, Leizhao Shen, Qiu Guan, Xiaoxin Li, Qianwei Zhou and Su Ruan},
pdfkeywords={Semantic segmentation, Multi-scale features, Attention mechanism, Feature fusion, COVID-19},
}

\begin{document}

\maketitle

\begin{abstract}
Due to the irregular shapes,various sizes and indistinguishable boundaries between the normal and infected tissues, it is still a challenging task to accurately segment the infected lesions of COVID-19 on CT images. In this paper, a novel segmentation scheme is proposed for the infections of COVID-19 by enhancing supervised information and fusing multi-scale feature maps of different levels based on the encoder-decoder architecture. To this end, a deep collaborative supervision (Co-supervision) scheme is proposed to guide the network learning the features of edges and semantics. More specifically, an Edge Supervised Module (ESM) is firstly designed to highlight low-level boundary features by incorporating the edge supervised information into the initial stage of down-sampling. Meanwhile, an Auxiliary Semantic Supervised Module (ASSM) is proposed to strengthen high-level semantic information by integrating mask supervised information into the later stage. Then an Attention Fusion Module (AFM) is developed to fuse multiple scale feature maps of different levels by using an attention mechanism to reduce the semantic gaps between high-level and low-level feature maps. Finally, the effectiveness of the proposed scheme is demonstrated on four various COVID-19 CT datasets. The results show that the proposed three modules are all promising. Based on the baseline (ResUnet), using ESM, ASSM, or AFM alone can respectively increase Dice metric by 1.12\%, 1.95\%,1.63\% in our dataset, while the integration by incorporating three models together can rise 3.97\%. Compared with the existing approaches in various datasets, the proposed method can obtain better segmentation performance in some main metrics, and can achieve the best generalization and comprehensive performance.The code is publicly available at \href{https://github.com/HuHaigen/COVID-19-Lung-Infection-Segentation}{https://github.com/HuHaigen/COVID-19-Lung-Infection-Segentation} or \href{https://github.com/slz674763180/COVID19}{https://github.com/slz674763180/COVID19}. The package includes the proposed three modules and joint loss function for reproducibility purposes.
\end{abstract}


\section{Introduction}
Since the outbreak of COVID-19 in December, 2019, it has spread rapidly around the world, and has caused millions of casualties and amount of economic losses. Rapid diagnosis of COVID-19 is of great significance for diagnosis, assessment and staging COVID-19 infection \cite{1,2,He2021}. Nucleic acid testing is the ``gold standard'' for the diagnosis of COVID-19, but the diagnosis are easily influenced by the quality of the sample collection, and it is also more time consuming. Therefore, it is still common to use the imaging diagnosis methods such as CT and Xray. Especially, the combining of artificial intelligence (AI) with other methods has been proposed to help auxiliary diagnosis by using medical images for COVID-19 in clinical practice, and some deep learning-based methods are becoming hot spots in the detection and segmentation of COVID-19 infected areas. For example, a modified inception neural network was proposed to train the Regions of Interest (RoI) instead of the whole CT images for classifying COVID-19 patients from control group \cite{23}. Amyar et al. \cite{Amyar2020} proposed a multitask deep learning model to jointly identify COVID-19 patient and segment COVID-19 lesion from chest CT images. Oulefki et al. \cite{Adel2021} presented the utility of an automated tool of segmentation and measurement for COVID-19 lung Infection using chest CT imagery. Owing to the fact that lung infected region segmentation is a necessary initial step for lung image analysis, some image segmentation algorithms are also proposed for some specific application scenarios. For instance, an improved Inf-Net was proposed to segment the infection area of the novel coronavirus, and a semi-supervised training method is put forward to solve insufficient amount of labeled CT and improve the segmentation performance \cite{16}. Currently, most of the methods are based on detection and classification tasks, but not much on the semantic segmentation of infection on CT slices \cite{Rorat2021}, so that the assessment and staging COVID-19 infection are greatly limited. Therefore, according to CT imaging characteristics, it is necessary to propose some segmentation methods for the infection regions of COVID-19, so that we can further achieve quantitative analysis of the lesions.\par
\begin{figure}[ht]
	\centering
	\subfigure[]{
	\label{fig_1a} 
	\includegraphics[width=0.23\textwidth]{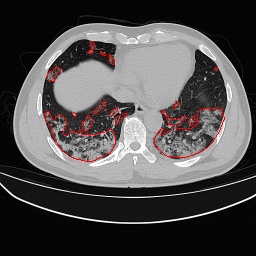}}
	\subfigure[]{
	\label{fig_1b} 
	\includegraphics[width=0.23\textwidth]{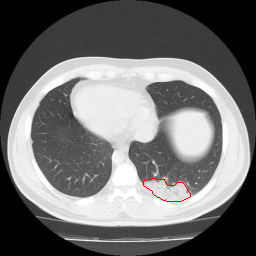}}
	\caption{An illustration of challenging task for identification the infected lesions (contours in red) of COVID-19 on CT images. (a) The infections have various scales and shapes.(b)There is no obvious difference between normal and infected tissues.}
	\label{fig:subfig} 
	\end{figure}

However, it is a still challenging task to accurately segment the infected lesions of COVID-19 on CT images owing to the following facts.
\begin{enumerate}
	\item The infections have irregular boundary, different sizes and shapes from slice to slice on CT images (shown in Figure \ref{fig_1a}). It would easily lead to missing some small ground-glass lesions or generating excessive over-segmentation for the infections on CT images.
	\item  There seems to be no discernible difference between infections and normal tissues (shown in Figure \ref{fig_1b}). It is unaffected for the detection or classification, but it can decrease segmentation accuracy and quantified quality.
	\item  The existing semantic segmentation approaches like the encoder-decoder structure exist a ``semantic gap'' between low-level visual features and high-level semantic concepts, which greatly limits the efficiency of semantic segmentation. 
	\end{enumerate}

To address these issues, a novel segmentation scheme is proposed for the infections of COVID-19 based on the encoder-decoder architecture \cite{7} in this paper, and the proposed scheme can collaboratively enhance supervised information of different levels and fuse different scale feature maps. For the proposed deep collaborative supervision scheme, we propose an Auxiliary Semantic Supervised Module (ASSM) and an Edge Supervised Module (ESM) to guide the network learning the features of edges and semantics in the encoding stage, respectively. As for multi-scale feature maps, an Attention Fusion Module (AFM), following with the decoding stage, is proposed to reduce the semantic gaps between high-level and low-level feature maps. The proposed attention fusion strategy can take full advantage of different scale context information. Finally, a series of experiments are conducted on the COVID-19 dataset to verify the effectiveness of the proposed scheme. The results show that our method can obtain better performance for the segmentation of COVID-19 infections than the existing approaches. The main contributions of this paper are listed as follows.
\begin{itemize}
\item{An ESM is put forward to highlight low-level boundary features. The edge supervised information is incorporated into the initial stage of down-sampling, as the proposed edge supervised loss function allows to capture rich spatial information in various scales.}
\item{An ASSM i s proposed to enhance high-level semantics from feature maps with different scales. The mask supervised information is introduced into the later stage of down-sampling, thanks to the corresponding auxiliary semantic loss function that is defined to explore sufficient semantic information from various scale infections on COVID-19 CT images.}
\item{An AFM is developed to fuse various scale feature maps from the up-sampling stage. An attention mechanism is utilized to reduce the semantic gaps between high-level and low-level feature maps, so as to strengthen and supplement the lost detailed information in high-level representations.}
\item{A joint loss function is constructed by combining the edge supervised loss, auxiliary semantic supervised loss and fusion loss. It can guide the network achieving a deep collaborative supervision on edges and semantics, and prompting the fusion efficiency on multiple scale feature maps from different levels.}
\end{itemize}

This paper is organized as follows. Section 2 introduces the related works. Section 3 describes details about the proposed methods, including Edge Supervised Module (ESM), Auxiliary Semantic Supervised Module (ASSM) and Attention Fusion Module (AFM). Section 4 presents experiments, results and discussions, and Section 5 concludes this work.

\section{Related works}
In this section, we provide a short review of previous studies on network models, edge supervision, multi-scale object recognition, and attention mechanism.
\label{sec:headings}

\subsection{Network Models}
Deep network models are a kind of hierarchical feature learning methods by learning multiple levels of representation to model complex relationships among data, and higher-level features and concepts are thus defined in terms of lower-level ones, and such a hierarchy of features is called a deep architecture \cite{Deng2013}. Usually, the first layers will learn the low level features like intensity, color, lines, dots and curves, then the more the layers approach the output layer, the more the layers will learn the high level features like objects and shapes in a feature extracting pipeline. For example, from AlexNet\cite{8}, VGG\cite{9} to ResNet\cite{10}, the ability of feature extraction is becoming more and more powerful with the deepening of the network depth. Accordingly, the deeper networks can provide a powerful feature extraction ability for semantic segmentation tasks, and can greatly improve segmentation accuracy. \par

Since FCN \cite{6} is proposed, other semantic segmentation networks attempt to improve this architecture by adding new modules to solve the problems regarding the lack of spatial and contextual information. For example, U-Net\cite{7} is greatly improved only by adding the skip connection based on FCN. PSPNet\cite{11} employs pyramid pooling module to explore the global context information, and it can improve the accuracy of target segmentation at different scales. Besides, DeepLabV3+\cite{12} combines the advantages of Spatial Pyramid Pooling (SPP) module and encoder-decoder structure, and further explore the Xception model and apply the depthwise separable convolution to both Atrous Spatial Pyramid Pooling (ASPP) and decoder modules. PSANet\cite{13} can capture pixel level relationship and relative position information in spatial dimension through convolution layer. In addition, EncNet\cite{14} also introduced a channel attention mechanism to capture the global context. \par

Although many advanced network structures have been emerged for semantic segmentation tasks, U-Net and its derivatives are still the most popular architecture and have been widely applied in the medical imaging community\cite{Hu2020,Hu2021}. However, despite their outstanding overall performance in segmenting medical images, the U-Net-based architecture seems to be lacking in certain aspects. For example, although the high-level feature map can be optimized through the concatenation the feature maps of the low-level layers and the high-level layer by using skipping connection, it is still very difficult to reduce the semantic gap between low-level visual features and high-level semantic features. Thus, we select ResUNet as the backbone to attempt to exploit a novel segmentation architecture for the COVID-19 segmentation task in this work. \par

\subsection{Edge Supervision and Multi-scale Object Recognition}
Edge information, as an important image feature, is drawing more and more attention in deep learning community owing to the fact that edge information is conducive to the extraction of object contour in segmentation tasks. For example, explicit edge-attention are utilized to model the boundaries and enhance the representations in \cite{16}. Wu et al.\cite{28} proposed a novel edge aware salient object detection method, and it passes messages between two tasks in two directions, and refines multi-level edge and segmentation features. ET-Net\cite{17} integrates edge detection and object segmentation into a deep learning network, and the edge attention representation is embedded to supervise the segmentation prediction. Normally, edge information can provide useful fine-grained constraints to guide feature extraction in semantic segmentation tasks. However,  high-level feature maps have little edge information, while low-level layers contain richer object boundaries.  \par

For the multi-scale object recognition problem, it is common practice to exploit multiple levels of coarse and fine-grained semantic features by adopting different network structures in computer vision. For example, the operations of convolution and pooling on the original image is used to obtain feature maps of different sizes, and it is similar to constructing pyramids in the feature space of images. Feature Pyramid Networks (FPN) \cite{Lin2017FeaturePN} is one of the most typical examples, and it adopts a top-down architecture with lateral connections for building high-level semantic feature maps at all scales. It has been demonstrated a significant improvement as a generic feature extractor in detection tasks, and has been widely applied in different detection architectures, such as Faster R-CNN \cite{Ren2015faster} and Mask R-CNN \cite{He2017mask}.\par

It is widely known that the low-level feature maps pay more attention to detail information, while the high-level lay much attention to semantic information. More specifically, 
the encoded pathway is mainly used for feature extraction, and there are hierarchy and gradation for various feature. Because the spatial resolution and the semantics can be decreased and strengthened along with the deepening of down-sampling, respectively. Significantly, FPN \cite{Lin2017FeaturePN} and U-Net  \cite{7} both adopt encoder-decoder architecture, but they are respectively applied in object detection and semantic segmentation. The main difference is that there are multiple prediction layers for various scale features in FPN \cite{Lin2017FeaturePN}. Inspired by this, we attempt to exploit sufficient multi-scale context information from different levels of the encoder in this work. Low level detailed feature maps can exploit rich spatial information, and they could strengthen the boundaries of the infected regions; while high-level semantic feature maps can endow position information, and they could locate the infected regions.\par

\subsection{Attention Mechanism}
Attention can be regarded as a mechanism, and it emphasizes the features that need attention through the context of feature maps. Normally, an attention mechanism is used to highlight the important context in the channel-wise or space-wise \cite{23, 24}, while suppressing the context information irrelevant to the content. For example, Fu et al. \cite{19} proposed Dual Attention Network (DAN), and two attention modules were introduced to capture the spatial dependence between any two positions in the feature maps. A similar self-attention mechanism was used to capture the channel dependence between any two channels, and the weighted sum of all channel was utilized to update each channel. Huang et al.\cite{20} proposed Criss-Cross Net (CCNet) to capture this important information in a more effective way, specifically, for each pixel, CCNet can obtain the context information on its crisscross path through a Criss-Cross attention module. Non-local operations, proposed by Wang et al.\cite{22}, can directly capture remote dependencies by calculating the interaction between any two locations. Besides, an attention mechanism is also used to aggregate different levels of features to bridge the semantic gaps between low-level features and high-level semantics. For example, Li et al.\cite{21} proposed Gated Fully Fusion(GFF)  to fully fuse multi-level feature maps controlled by learned gate maps, and the novel module can bridge the gap between high resolution with low semantics and low resolution with high semantics. Inspired by this, we adopt an attention mechanism to fuse various level feature maps, and the proposed AFM can reduce the semantic gaps between high-level and low-level feature maps, so as to strengthen and supplement the missing detailed information in high-level representations.\par

\section{Methods}
\label{sec:others}

In this section,  we first present the proposed network architecture. Then we introduce in details the proposed three modules: ESM, ASSM and AFM. 
\subsection{Proposed Network Architecture}
\begin{figure*}[htbp]
\begin{center}
\includegraphics[width=0.75\textwidth]{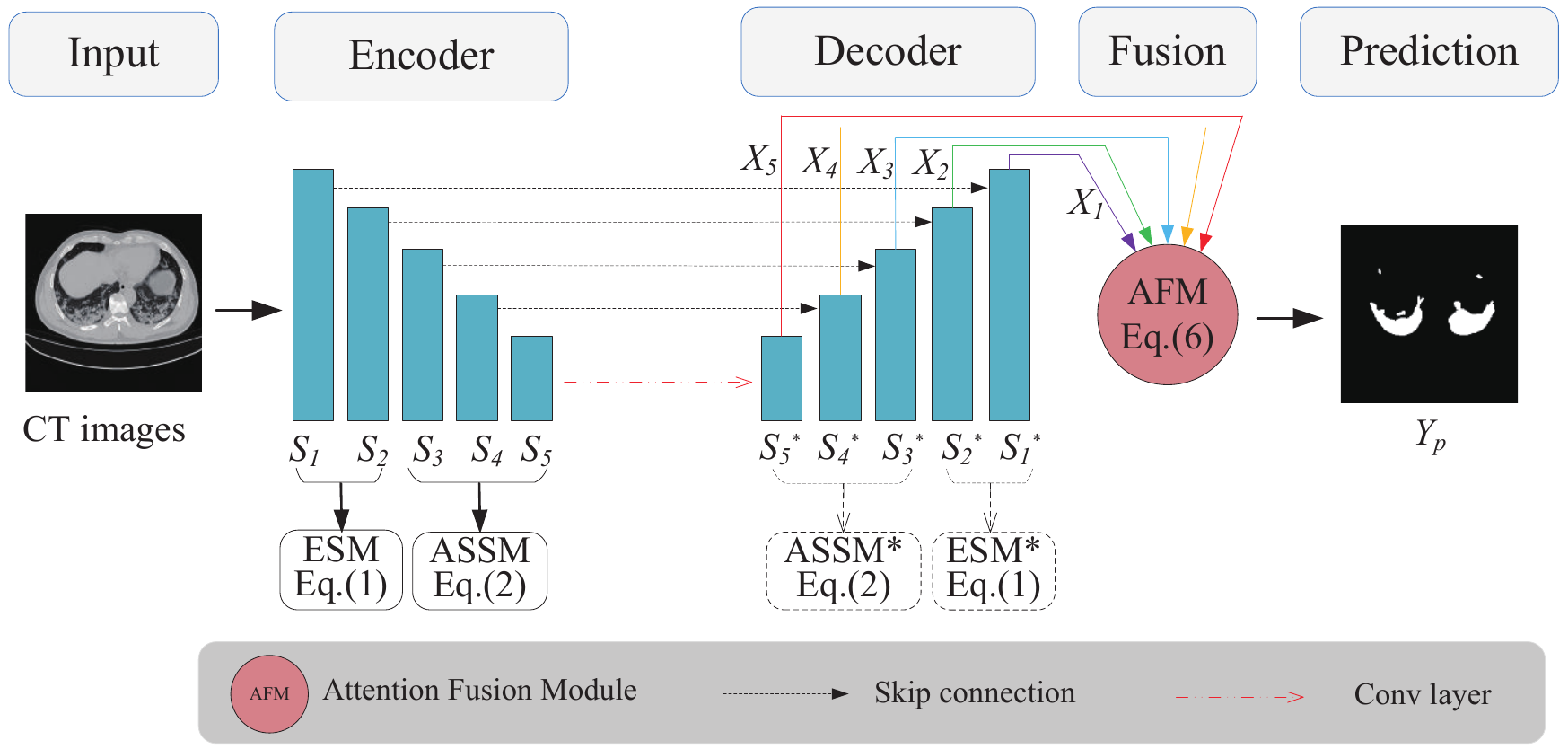}
\caption{An Illustration of the overall network architecture. The proposed architecture comprises of ASSM, ESM and AFM based on encoder-decoder structure. (1) ESM is used to  further highlight the low-level features in the initial shallow layers of the encoder, and it can capture more detailed information like object boundaries. (2) While ASSM is employed to strengthen high-level semantic information by integrating object mask supervised information into the later stages of the encoder. (3) Finally, AFM is utilized to fuse multi-scale feature maps of different levels in the decoder.}
\label{fig_2}
\end{center}
\end{figure*}
As mentioned above, U-Net \cite{7} and FPN \cite{Lin2017FeaturePN} both have a similar encoder-decoder structure for multi-scale object vision tasks, consisting of a contracting path to capture context and a symmetric expanding path that enables precise localization. While U-Net \cite{7} creates a path for information propagation allowing signals propagate between low and high levels by copying low level features to the corresponding high levels. Despite achieving good segmentation performance in U-Net and its variations, however, the edge information and channels would decrease and increase along with down-sampling of the contracting path, respectively. Both cases can lead to effective information missing, thereby not exploring sufficient information from full scales so as to suffer segmentation performance degradation. While FPN \cite{Lin2017FeaturePN} can overcome these drawbacks to retain multi-scale contextual information by using multiple prediction layers: one for each up-sampling layer. Based on this idea, we propose a novel segmentation scheme for the infections of COVID-19.  \par

Figure \ref{fig_2} illustrates the proposed network architecture. Firstly, we collaboratively enhance the supervised information by introducing edge and semantic information into the encoding stage. Note that the initial stages are used for the edge supervision, while the later stages for the semantic supervision. They occupy the whole down-sampling together, more precisely, the sum of the low-level and high-level layers is equal to the total layers of the encoder. Especially, low-level feature maps from shallow layers are with high resolution, but with limited semantics, whereas high-level feature maps from deep layers have low spatial resolution without detailed information (like object boundaries). When various levels are selected to enhance the supervised information, there is a trade-off between edge supervision and semantic supervision, thus we call it ``collaborative supervision'' (``Co-supervision'').  Then we fuse multi-scale feature maps of different levels from the decoding stage in an encoder-decoder framework (like U-Net). Considering the fact that low level detailed feature maps have high resolution and can capture rich spatial information like object boundaries, we design an ESM to highlight low-level boundary features by incorporating the edge supervised information into the initial stage (like $S_1$ and $S_2$ in Figure \ref{fig_2}) of down-sampling in the encoder. While high-level semantic feature maps embody position information like object concepts, thus we present an ASSM to strengthen high-level semantic information by integrating object mask supervised information into the later stage (like $S_3\sim S_5$ in Figure \ref{fig_2}). Finally, the obtained various scale feature maps from the up-sampling stage are fused by adopting an attention mechanism to achieve good segmentation performance for infections of COVID-19.\par
\begin{figure}[ht]
\centering
\subfigure[]{
\label{fig_3a} 
\includegraphics[width=0.3\textwidth]{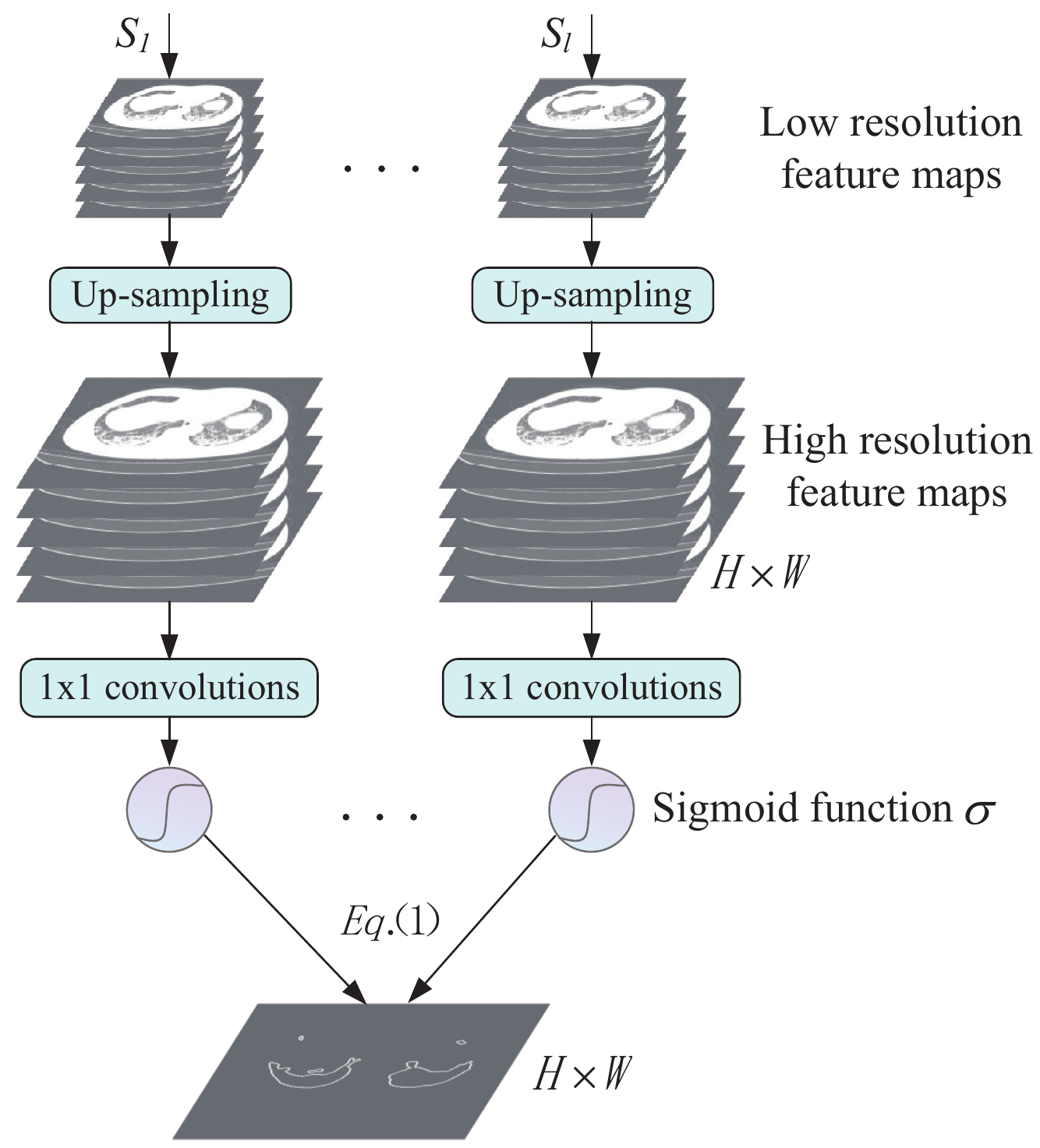}}
\hspace{0.5 in} 
\subfigure[]{
\label{fig_3b} 
\includegraphics[width=0.3\textwidth]{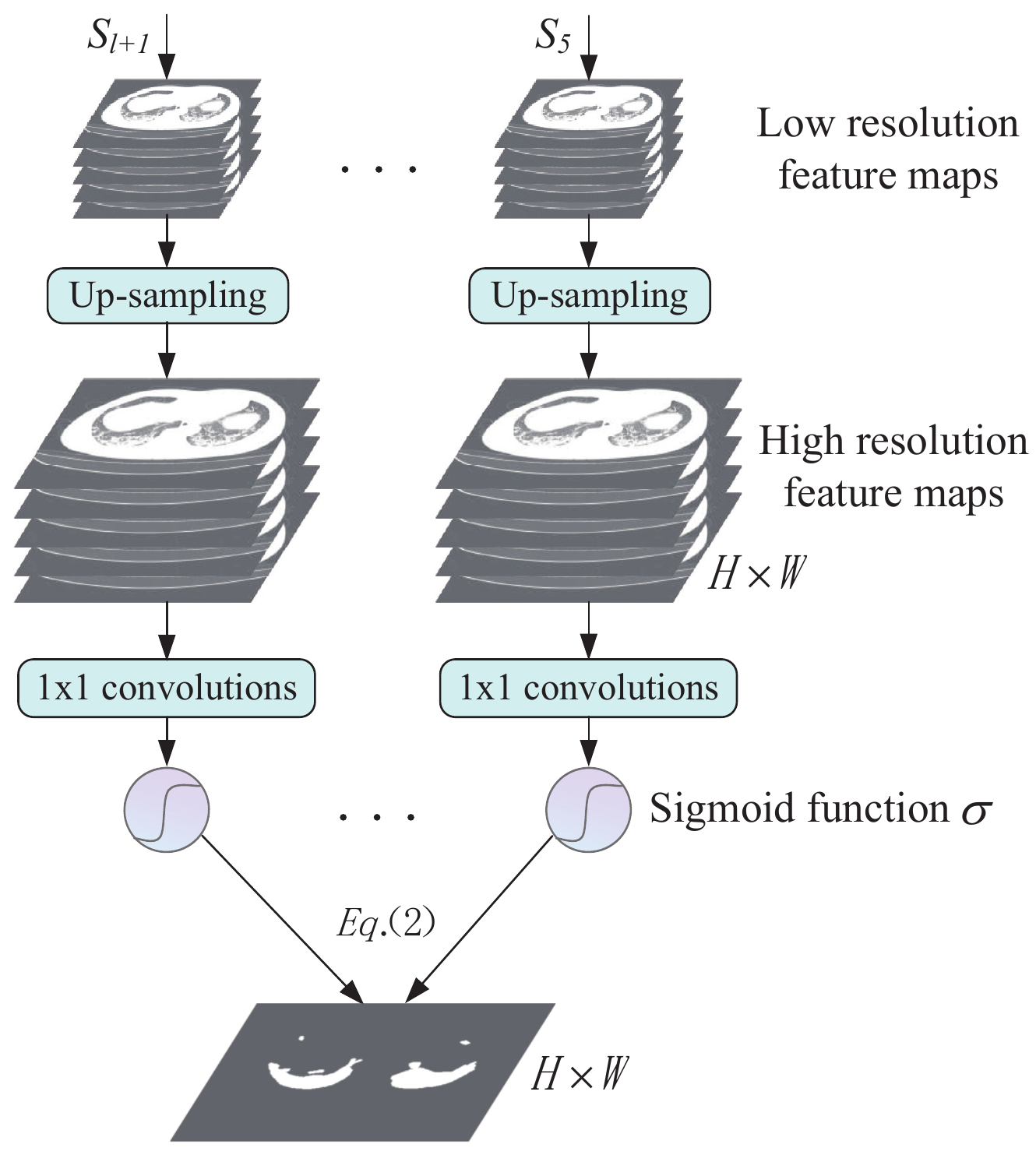}}
\caption{An illustration of ESM and ASSM. Firstly, the low resolution feature maps from the stage $S_i$ are resized to the same size $H\times W$ with the input image by using bilinear interpolation up-sampling. Then all high resolution feature maps are reduced to a feature map by using $1\times 1$ convolutions. Finally each pixel value of the obtained feature map is converted to a probability by using  Sigmoid function $\sigma(\cdot)$, and the prediction image of the $S_i$ stage is obtained. (a) ESM: the edge supervision is achieved by comparing between the obtained edge prediction image $S_{edge}^i$ and the corresponding edge Ground Truth (GT) $G_{edge}$ based on $Eq.(1)$. (b)ASSM: the auxiliary semantic supervision is achieved by comparing between the obtained coarse segmented image $S_{mask}^i$ and the corresponding Ground Truth (GT) of segmentation mask $G_{mask}$ based on $Eq.(2)$.}
\label{fig:subfig} 
\end{figure}
\subsection{Edge Supervised Module (ESM)}
Many studies \cite{28,17} show that the edge information can provide effective constraints to the feature extraction in the segmentation task. To supplement the missing edge information along with down-sampling, we propose ESM to further highlight the object boundary features in the low-level layers. Because feature maps of low level from shallow layers are with high resolution and detailed information (including edge information), and these detailed information are easily lost during the initial stage of the down-sampling process, the proposed ESM can capture more detailed information like object boundaries. Specifically, we can guide the network to extract edge features from the initial stages like $S_1$ and $S_2$ (shown in Figure \ref{fig_2}) by defining edge supervised loss function. To this end, the output feature maps from the initial stage are firstly resized to the size $H\times W$ of the original image by using bilinear interpolation up-sampling. Then the obtained large feature maps of each layer in ESM are reduced to a feature map by using $1\times 1$ convolution operation. Finally each pixel value is converted to a probability by using Sigmoid function $\sigma(\cdot)$  (shown in Figure \ref{fig_3a}), and an edge prediction image with $H\times W$ is obtained. Accordingly, the edge supervised loss function is given based on Dice coefficient as follows. 
\begin{equation}
L_{edge} = 1 - 2\times\sum_{i=1}^l\frac{\zeta_i({S_{edge}^i\cap{G_{edge}}})}{(S_{edge}^i+G_{edge})}
\end{equation}
where $S_{edge}^i$ is the edge prediction image obtained by using bilinear interpolation up-sampling in the $i^{th}$ stage. $G_{edge}$ is the corresponding Ground Truth (GT) of edge image, which is obtained by generating edge GT from the segmentation mask. $l$ is the number of stages used for edge supervised in the ESM. $\zeta_i$ ($i=1, \dots, l$) is the weight coefficient of the $i^{th}$ stage. By using skip connections and AFM, the edge features in the high-level feature maps can also be strengthened. 
\subsection{Auxiliary Semantic Supervised Module (ASSM)}
For the multi-scale object segmentation, the multi-level loss function is used to build receptive fields of different sizes for different layers in the network. For example, FPN \cite{Lin2017FeaturePN} uses multi-level auxiliary loss to detect objects at different scales, and it is a great breakthrough in multi-scale object detection task. Inspired by this, we develop an ASSM based on the similar strategy in our network. Specifically, the semantic information is gradually strengthened along with the down-sampling process in the encoder, and the high-level feature map has  rich semantics but low spatial resolution without detailed information. Different layers contain different level semantic features according to the feature hierarchy of the contracting path. Thereby we can define an auxiliary semantic loss function to reduce the semantic gaps between high-level and low-level feature maps in the later stage (i.e.,$S_3\sim S_5$) of the encoder. Eventually, low-level semantic features can be strengthened by using multi-scale skip connections and AFM, and it can also reduce the background noise in the low-level feature maps.\par
Similar to the above steps in ESM, we can obtain one coarse segmented image with the size of $H\times W$ and the probability of each pixel through a series of operations, such as bilinear interpolation, $1\times 1$ convolution, and Sigmoid function $\sigma(\cdot)$ (shown in Figure \ref{fig_3b}). Then the auxiliary semantic loss function is defined based on Dice coefficient as follows.   
\begin{equation}
L_{semantic} = 1 - 2\times\sum_{i=l+1}^5\frac{\omega_i({S_{mask}^i\cap{G_{mask}}})}{(S_{mask}^i+G_{mask})}
\end{equation}
where $S_{mask}^i$ and $G_{mask}$ are the obtained coarse segmented image in the $i^{th}$ stage of the Encoder and the Ground Truth (GT) of segmentation mask, respectively. $\omega_i$ ($i=l+1, \dots, 5$) is the weight coefficient of the $i^{th}$ stage.
\subsection{Attention Fusion Module (AFM)}
As mentioned above, high-level features are very efficient in semantic segmentation tasks. However, the high-level feature maps easily lead to inferior results for small or thin objects owing to the fact that the operations of convolution and pooling can cause the detailed information missing, thereby high-level feature maps have coarse resolution. To compensate the lost detailed information in high-level representations, it is necessary to import low level features. However, the full-scale skip connections can only incorporate low-level details with high-level semantics from feature maps in different scales of the same level, and the semantic gaps existing among various levels hampers the effectiveness of the semantic segmentation. Thus we propose the AFM to fuse multi-scale feature maps of different levels by using an attention mechanism to strengthen and supplement the lost detailed information in high-level representations. \par
\begin{figure}[htbp]
\centering
\includegraphics[width=0.48\textwidth]{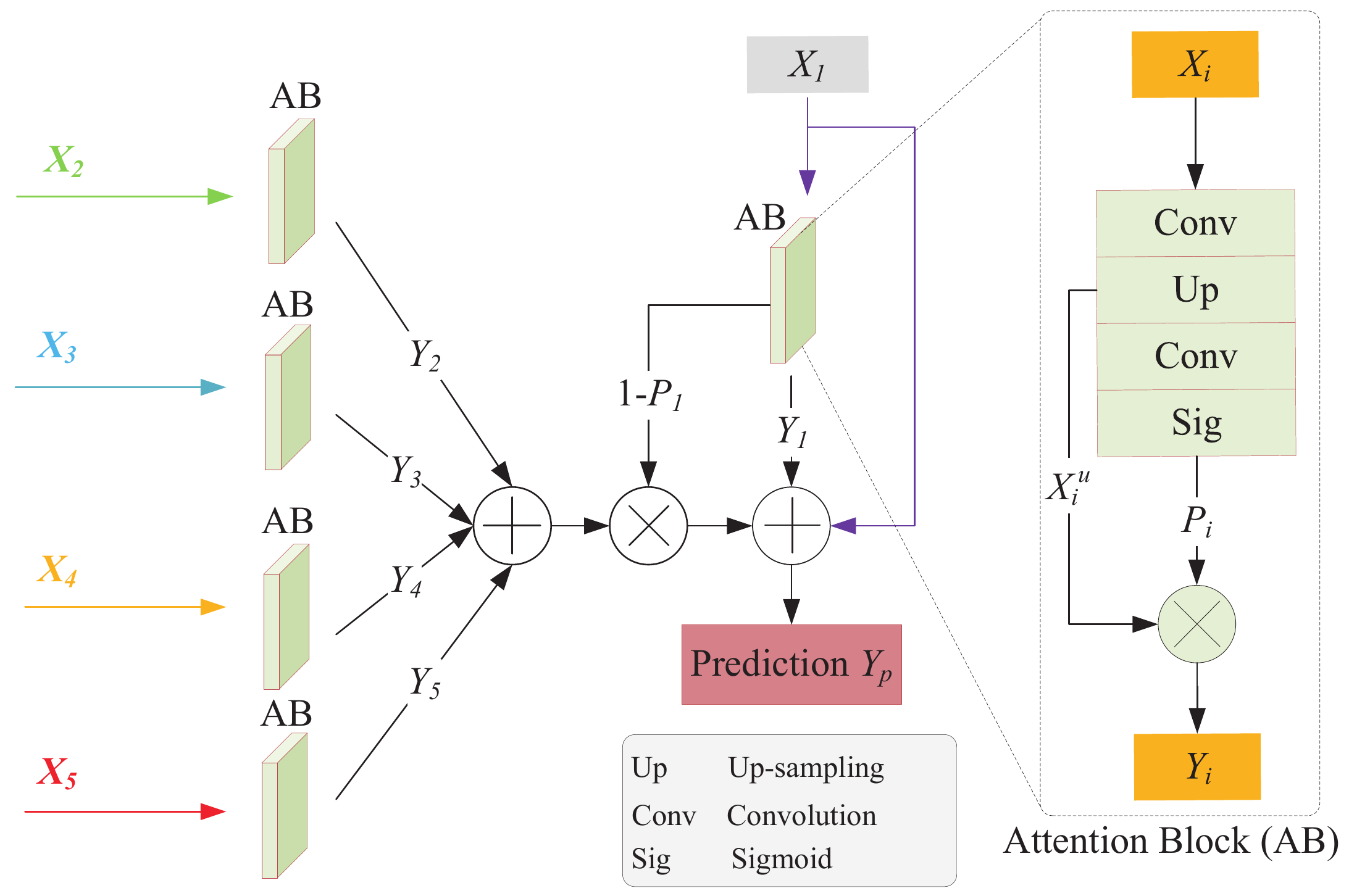}
\caption{An illustration of the attention mechanism. $X_i^{u}$ represents the up-sampling intermediate result by bilinear interpolation for the feature map $X_i$, and its 2D size is the same size with the input image.}
\label{fig_4}
\end{figure}
Gated Fully Fusion(GFF) \cite{21} can selectively fuse features from multiple levels using gates in a fully connected way, and add weights to each spatial position by using skip connection. Inspired by this idea, an attention mechanism is incorporated into the AFM by aggregating different level features, aiming at reducing the semantic gaps between low-level features and high-level features. The corresponding attention mechanism is illustrated in Figure \ref{fig_4}. In general, we can directly obtain the segmentation maps from the top feature map $X_1$($\in R^{C\times{H}\times{W}}$, where $c$, $h$ and $w$ are the channel number, height and width, respectively) of the expansive path in the standard U-Net. The $X_1$ has high spatial resolution because the outputs need to be with the same resolution as the input image, but actually, multiple down-sampling and up-sampling operations make the deep network cause mistake and loss in the detailed information. As well as strengthening the top feature map $X_1$, therefore, we can aggregate feature maps of other levels (i.e., $X_2\sim X_5$) to supplement the lost detailed information caused by the filters or pooling operations.  \par
\begin{figure*}[htbp]
\centering
\includegraphics[width=0.75\textwidth]{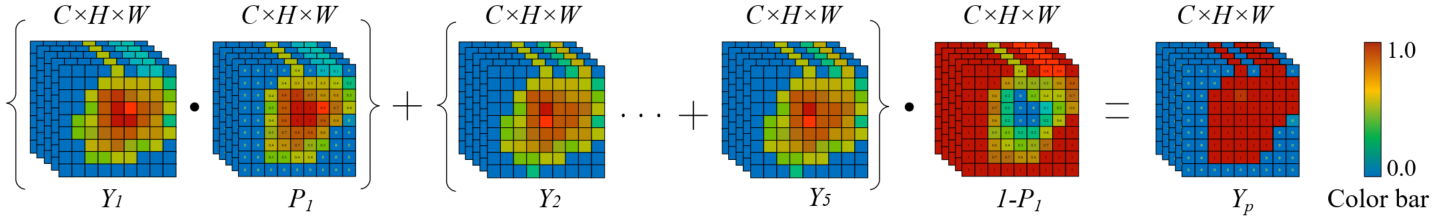}
\caption{The procedure of the attention block. The color bar represents the trends of confidence values, and the red and blue denote 1 and 0, respectively.}
\label{fig_5}
\end{figure*}
More precisely, we can obtain a confidence map $P_1$($\in R^{C\times{H}\times{W}}$) through the attention block (AB) of the top feature map $X_1$. The points with high confidence have a greater possibility to retain the original feature map values, and vice versa. Similarly, the lost detailed information is represented by the confidence map $1-P_1$, in which the higher the value, the less object information it contains. Thus, we can strengthen the top feature map $X_1$ through the dot product between the confidence map $P_1$ and $X_1$, and can supplement the lost detailed information by using dot product between the confidence map $1-P_1$ and the sum of other feature maps. The procedure of the attention block is  illustrated in  Figure \ref{fig_5}, and the final prediction result $S_P$ can be defined as follows.
\begin{equation}
S_P = X_1 + Y_1 + (1-P_1) \cdot \sum_{i=2}^{5}Y_i
\end{equation}
where $Y_i$ is the output by using the attention block to process the corresponding $X_i$. While $X_i$ is firstly up-sampled to the same size with the input image by bilinear interpolation. Then $Y_i$ can be obtained by processing the up-sampling intermediate result $X_i^{u}$ based on the attention block, and it is defined as follows. 
\begin{equation}
Y_i = \Phi_A (X_{i})=P_i \cdot X_i^{u} 
\end{equation}
where $\Phi_A(\cdot)$ is the attention function.\par
The specific process is as follows. 
\begin{enumerate}
\item{Each up-sampling feature map ($X_i$) is processed through an attention block. }
\item{After an $1\times 1$ convolution operation, the channels are reduced to $64$, and we can obtain the $i^{th}$ level feature maps. }
\item{Then the resolution is resized to $H\times W$ by using bilinear interpolation.} 
\item{After the operation of a convolution and Sigmoid function $\sigma(\cdot)$, we can obtain the confidence output $Y_i$ by using a dot product  $Y_i=P_i \cdot X_i^{u}$. Note that the top feature map $X_1$ is selected as the main prediction, while other confidence output only as the supplement of $Y_1 = \Phi_A(X_1)$. When $P_1$ is small, it means that the corresponding confidence is low, and thereby we can compensate the lost information by doing a dot product between ($1-P_1$) and the sum of the confidence outputs of other layer feature maps $X_i$ ($i=2,\dots,5$). }
\item{Finally, the final prediction result $S_p$ is obtained by summing the residuals of $X_1$. The specific process is shown in Algorithm $\ref{A1}$. } 
\end{enumerate}
The loss function for fusion is defined as follows:

\begin{equation}
L_{fusion} = 1 - \frac{2\times({S_p\cap{G}})}{S_p+G}
\end{equation}
where $G$ represents the ground truth of COVID-19.
\begin{algorithm}
\caption{Fusion Algorithm}
\label{A1}
\begin{algorithmic}[1]
\Require Feature map while up-sampling $X_i$ (i$\in$[1,\dots,5])
\Ensure Prediction $S_p$
\State Adopt $1\times 1$ convolution on $X_i$ to change its channel number to 64 
\State Resize the above obtained feature maps to the original image size of $H\times W$ by using up-sample, and obtain $X_i^{u}$
\State Adjust $X_i^{u}$ to one channel by using $3\times 3$ convolution 
\State Generate the confidence map $P_{i}$ by adopting Sigmoid function $\sigma(\cdot)$
\State Obtain $Y_{i}$ by doing a dot product between $X_i^{u}$ and $P_{i}$ in each channel, and perform the sums $\sum_{i=2}^{5}Y_i$
\State Do a dot product between ($1-P_1$) and $\sum_{i=2}^{5}Y_i$,  and obtain $(1-P_1)\sum_{i=2}^{5}Y_i$
\State Obtain the prediction $S_p$ by calculating the summation of $X_1$, $Y_1$ and $(1-P_1)\sum_{i=2}^{5}Y_i$
\State \Return{$S_p$} 
\end{algorithmic}
\end{algorithm} 
\begin{equation}
L_{total} =\theta L_{edge}+\beta L_{semantic}+L_{fusion}
\label{equ6}
\end{equation}
where $\theta$ and $\beta$ are weight coefficients.\par
Considering the fact that there would be negative values in the category imbalance case when using the cross-entropy loss function. Therefore, we select Dice loss to supervise the predictions and labels in our experiments. To achieve deep fusions and supervisions for the features of different level, the overall loss function integrates ESM, ASSM and AFM, given as Eq. \ref{equ6}.

\section{Experiments}
\subsection{Datasets and Baselines}
We collect the COVID-19 segmentation dataset from two sources. One is from \cite{29}, including more than 900 CT images, among them about 400 slices with infections. Another is from \cite{30}, and it contains 3D CT images of 20 patients, and we can obtain 3686 images by converting from 3D volumes into 2D slices. Due to the small datasets, the two sources are put together in a total of 4449 2D slices, among them 4000 slices for training sets and 449 slices for test sets, respectively. The GT contains four categories: $0\sim 3$ represent background, ground glass, consolidation and plural effect, respectively. Owing to the imbalance of infection categories in the dataset, for example, only few slices contain plural effect infection, we take all types of infection as one type. Considering the limitation of GPU memory, we resize the image resolution of $512\times 512$ to $256\times 256$ by bilinear interpolation, then Z-score is used for data normalization. Besides, to further verify the effectiveness and generalization ability of the proposed method, we select  three additional public COVID-19 datasets for testing and comparison, including MosMedData \cite{Morozov2020}, UESTC-COVID-19 \cite{Wang2020} and COVID-ChestCT \cite{Cohen2020}. MosMedData is a dataset of 100 axial CT images from more than 40 patients with COVID-19, including 829 slices with 512x512 size (see \cite{Morozov2020} for details),  and UESTC-COVID-19 contains CT scans (3D volumes) of 50 patients diagnosized with COVID-19 from 10 different hospitals(see \cite{Wang2020} for details). While COVID-ChestCT is a small dataset, and it contains 20 CT scans of patients diagnosed with COVID-19 as well as segmentations of lungs and infections made by experts (see \cite{Cohen2020} for details).
\begin{figure*}[htbp]
\begin{center}
\includegraphics[width=1\textwidth]{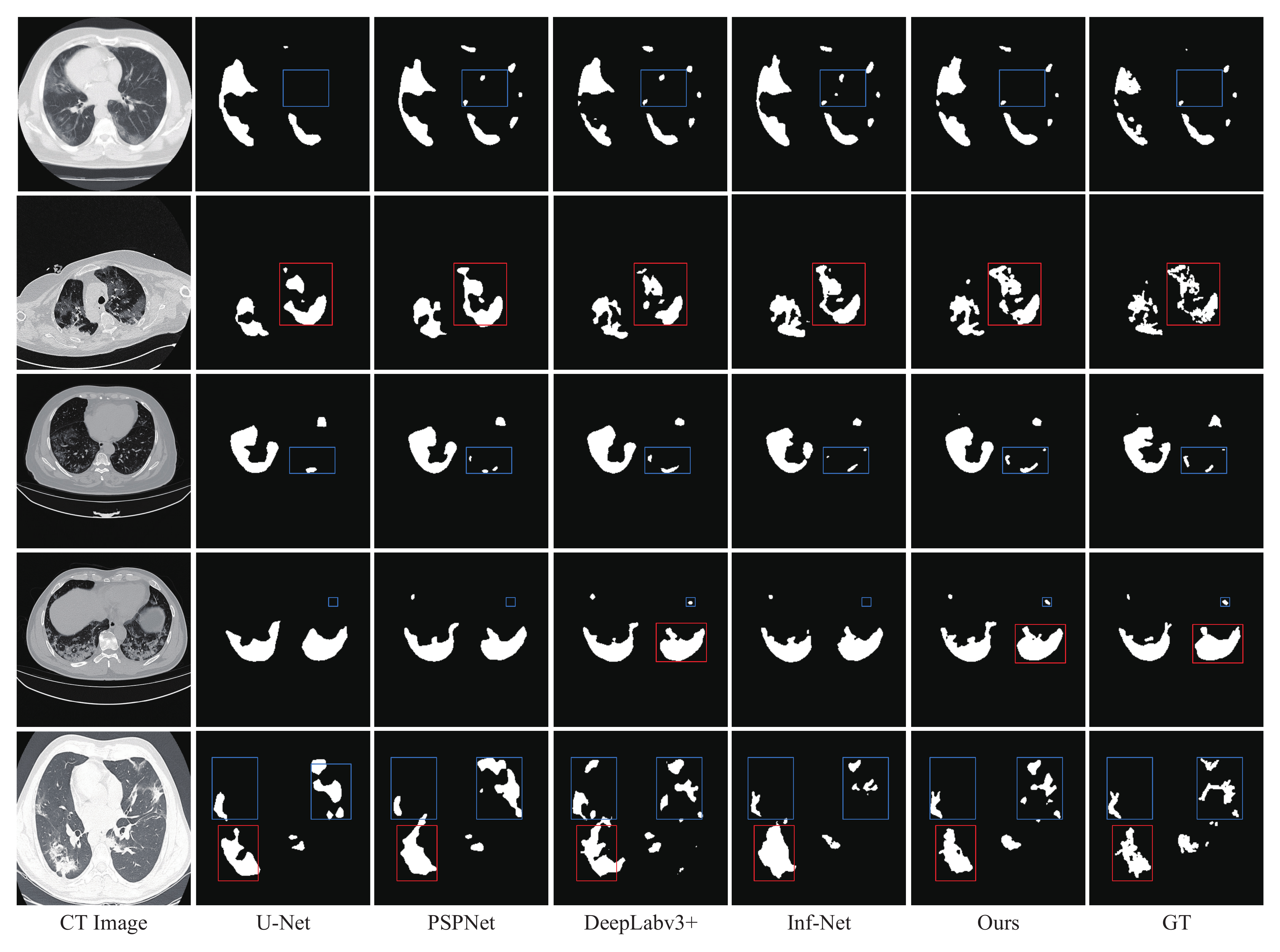}
\caption{Visual qualitative comparison of lung infection segmentation results among U-Net, PSPNet, DeepLabv3+, Inf-Net and the proposed method. Column 1:  the original CT image; Column 2:  U-Net; Column 3:  PSPNet; Column 4:  DeepLabv3+; Column 5:  Inf-Net; Column 6:  our method; Column 7:  the corresponding ground truth (GT).}
\label{fig_6}
\end{center}
\end{figure*}

\begin{figure*}[htbp]
\begin{center}
\includegraphics[width=1\textwidth]{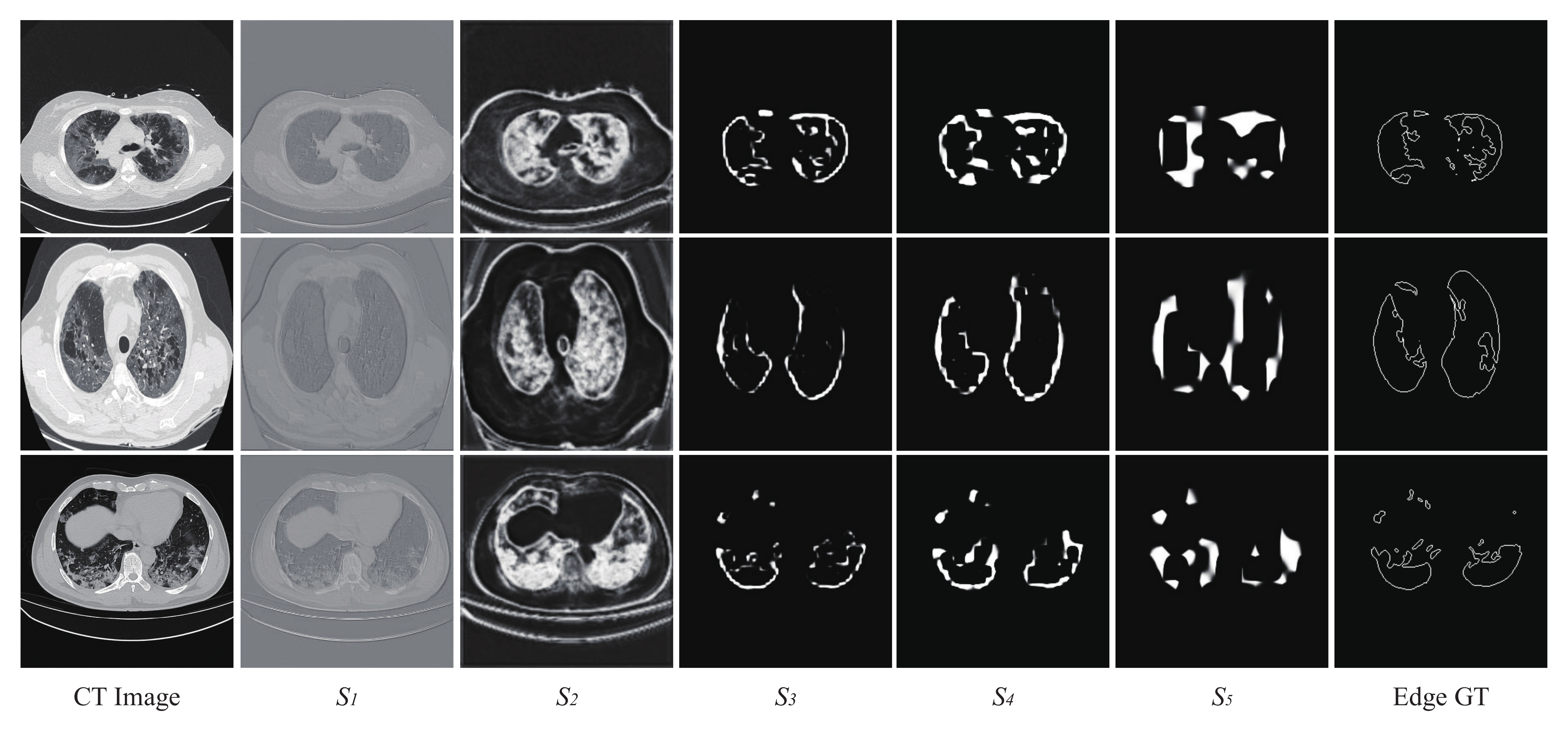}
\caption{Visualization of each stage supervised by ESM. Column 1: the original CT image; Columns 2 to 6: $S_1$ to $S_5$; Column 7: the corresponding edge ground truth (GT).}
\label{fig_7}
\end{center}
\end{figure*}

We select ResUNet as the backbone of the proposed network, in which the down-sampling of U-Net is replaced with ResNet. To verify the effectiveness of the proposed scheme, we use a series of popular segmentation models for comparison in the medical image segmentation area, such as U-Net\cite{7}, UNet++\cite{25}, and Attention U-Net \cite{33}, and we compare our methods with two cutting-edge models from the semantic segmentation: DeepLabV3+  \cite{12} and PSPNet \cite{11}.

\subsection{Evaluation Metrics and Experimental Settings}
We adopt three metrics to evaluate our methods, such as Dice similarity coefficient, Sensitivity (Sens.), Precision (Prec.). Besides, we also introduce three golden metrics to verify the detection and segmentation performance from the object detection field, such as Structure Measure\cite{26}, Enhance-alignment Measure\cite{27}, and Mean Absolute Error. In our evaluation, we select $S_p$ as the final output prediction, and measure the similarity/dissimilarity between $S_p$ and  ground-truth $G$, which can be formulated as follows. 
\begin{itemize}
\item{
Dice similarity coefficient: it is used to measure the proportion of intersection between $S_p$ and $G$, which is defined as follows.
\begin{equation}
Dice=\frac{2\times({S_p\cap{G}})}{S_p+G}
\end{equation}
}
\item{
Structure Measure (S$_{\alpha}$): it is used to measure the structural similarity between a prediction S$_p$ and ground-truth $G$, which is more consistent with the human visual system.
\begin{equation}
S_{\alpha} = (1 - \alpha) \times S_o(S_p, G) + \alpha \times S_r(S_p, G),
\end{equation}
where  $S_o$ and $S_r$ are the object-aware similarity and region-aware similarity, respectively.  $\alpha$ is a balance factor between $S_o$ and $S_r$. We report S$_{\alpha}$ using the default setting ($\alpha$ = 0.5) suggested in the original paper.
}
\item{
Sensitivity ($Sens.$): it is used to measure the percentage of positive samples in the total number of patients, or the probability of no missed diagnosis. The formulation is given as follows.
\begin{equation}
Sens.=\frac{{S_p\cap{G}}}{G}
\end{equation}
}
\item{
Precision ($Prec.$): it is used to measure the percentage of samples with negative test in the total number of healthy people, or the probability of not misdiagnosing. The formulation is given as follows:
\begin{equation}
Prec.=\frac{{S_p\cap{G}}}{S_p}
\end{equation}
}
\item{
Enhanced-alignment Measure (E$^{mean}_{\phi}$): it is a recently proposed metric for evaluating both local and global similarity between two binary maps. The formulation is given as follows:
\begin{equation}
E_{\phi}= \frac{1}{w\times{h}} \sum _{x}^{w}\sum _{y}^{h}{\phi}(S_p (x,y),G(x,y))
\end{equation}
where $w$ and $h$ are the width and height of ground-truth $G$, and $(x, y)$ denotes the coordinate of each pixel in $G$. Symbol $\phi$ is the enhanced alignment matrix. We obtain a set of $E_{\phi}$ by converting the prediction $S_p$ into a binary mask with a threshold from $0$ to $255$. In our experiments, we report the mean of E$_\xi$ computed from all the thresholds.
}
\item{
Mean Absolute Error ($MAE$): it is used to measure the pixel-wise error between $S_p$ and $G$, which is defined as:
\begin{equation}
MAE= \frac{1}{w\times{h}} \sum _{x}^{w}\sum _{y}^{h}|S_p (x,y)-G(x,y)|.
\end{equation}
}
\end{itemize}

For the hyper-parameters in the experiments is given in Table \ref{tab1}  by try-and-error, respectively. Note that the learning rate is initially selected as 1e-4, then is reduced by a factor of 0.5 when the test loss is not improved within 25 epoch. Early stopping is used to avoid over-fitting. All experiments are conducted on a desktop computer with an E3-1230 v5 3.40GHz 8-core processor, and with a GeForce GTX 1070 graphics card. A GPU implementation accelerates the forward propagation and back
propagation routines by using the Adam optimizer under the Pytorch framework. Each experiment is run three times, then its average and standard deviation $\pm$ are obtained.
\begin{table}[!htbp]
\caption{Hyperparameter setting}
\small
\centering
\begin{tabular}{ll}
\toprule[1pt]
Parameters           &   Values               \\   \midrule[0.7pt]
Input image size $H\times W$   &  $256\times 256$  \\   
batch\_size           &  8                         \\   
learning rate         &  1e-4                    \\    
Early stopping      &  25 epochs            \\    
$\theta$               &  0.8                      \\     
$\beta$                &  0.4                      \\     
$\zeta_i$ ($i=1, \dots, l$)       &  1    \\
$\omega_i$ ($i=l+1, \dots, 5$) &  1    \\
\bottomrule[1pt]
\end{tabular}
\label{tab1}
\end{table}
\subsection{Experimental Results}
\subsubsection{Quantitative results}
A series of  comparison experiments are implemented on our dataset, and the results are shown in Table $\ref{tab2}$. From Table $\ref{tab2}$, the proposed method can achieve the best performances among these methods in $Dice$, $Sens.$ and $Prec.$. Thereinto, our method has improved by around 4.4\% and 1.44\% in the main metric---Dice coefficient compared with U-Net \cite{7}  and Inf-Net \cite{16}, respectively. In particular, UNet++ \cite{25} and Attention U-Net \cite{33} represent the best U-Net-based methods in the medical image processing area, while Inf-Net \cite{16},  CE-Net \cite{Gu2019} and CPFNet \cite{Feng2020} are the newest and best methods for the segmentation of medical images. It suggests that the proposed scheme is effective and competitive, and can effectively fuse the multi-scale and multi-level features to accurately achieve the COVID-19 infection segmentation. \par
\begin{table*}[]
\begin{center}
\caption{Comparisons between different networks on our dataset. Bold black text and blue text represent the first and second best results, respectively.}
\small
\begin{tabular}{lllllll}
\toprule[1pt]   
Methods& $Dice(\%)\uparrow$  &$Sens.(\%)\uparrow$  &$Prec.(\%)\uparrow$  & $MAE(\%)\downarrow$  & $E_{\phi}$(\%)$\uparrow$  &$S_{\alpha}$(\%)$\uparrow$    \\ 
\midrule[1pt]   
U-Net \cite{7}                    & 85.56$\pm$0.33 & 85.38$\pm$1.53  & 85.76$\pm$0.89  & 0.72$\pm$0.01   & 94.21$\pm$0.02      & 81.23$\pm$0.23   \\
UNet++ \cite{25}               &86.71$\pm$1.25  &\textcolor{blue}{90.27$\pm$0.61}  &88.30$\pm$1.05  &0.60$\pm$0.02   &94.50$\pm$0.63       &84.61$\pm$1.00    \\
Attention U-Net\cite{33}   &87.40$\pm$0.26  &89.48$\pm$0.49  &89.88$\pm$0.53  &0.58$\pm$0.03   &94.74$\pm$0.87       &84.71$\pm$0.87    \\
PSPNet \cite{11}               & 87.45$\pm$0.31 & 88.32$\pm$1.25  & 89.89$\pm$1.11  & 0.60$\pm$0.05   & 93.84$\pm$0.27      & 83.81$\pm$0.20   \\
Deeplabv3 \cite{12}   & 87.81$\pm$0.19 & 89.24$\pm$0.96  &90.72$\pm$0.66 & 0.58$\pm$0.02&95.58$\pm$0.22& 86.03$\pm$0.95   \\
Inf-Net \cite{16}&     88.49$\pm$0.17  &90.07$\pm$0.35 & 90.39$\pm$0.18 & 0.55$\pm$0.01 &\textbf{95.70$\pm$0.24} &86.55$\pm$0.09   \\ 
SCRN \cite{28}                &86.24$\pm$0.08   &83.64$\pm$0.36  & 89.65$\pm$0.59  & 0.60$\pm$0.015   & 95.02$\pm$0.40      & 84.09$\pm$0.26   \\
F3Net \cite{Wei2020}       &87.99$\pm$1.45   &85.14$\pm$2.23  &\textcolor{blue}{91.08$\pm$0.17} & 0.58$\pm$0.025   & 93.51$\pm$0.63      &86.35$\pm$2.03  \\ 
DANet \cite{19}  & \textcolor{blue}{88.94$\pm$0.29} & 85.48$\pm$2.74  & 90.50$\pm$0.53  & 0.57$\pm$0.015   & 94.11$\pm$0.91    &86.90$\pm$1.33  \\ 
ACFNet \cite{Zhang2019}&83.25$\pm$0.18 & 83.88$\pm$0.10& 83.06$\pm$0.25&0.34$\pm$0.001& 85.21$\pm$0.15&90.62$\pm$0.05  \\ 
CE-Net \cite{Gu2019} &81.49$\pm$0.75 & 84.21$\pm$0.85  & 84.18$\pm$0.34  &\textbf{0.30$\pm$0.005} & 85.06$\pm$0.26    &\textcolor{blue}{92.00$\pm$0.10}  \\  
CPFNet \cite{Feng2020} & 85.19$\pm$0.14& 84.66$\pm$1.32  & 85.22$\pm$1.05  &\textcolor{blue}{0.31$\pm$0.002}& 86.38$\pm$0.07    &\textbf{92.09$\pm$0.20}  \\ \hline
ResUNet\_C$_2$F(\textbf{Ours}) & \textbf{89.93$\pm$0.09} & \textbf{90.29$\pm$0.66}  & \textbf{91.91$\pm$0.97}  &0.52$\pm$0.01 &\textcolor{blue}{95.69$\pm$0.10}  & 86.75$\pm$0.07 \\ 
\bottomrule[1pt]    
\end{tabular}
\label{tab2}
\end{center}
\end{table*}
Besides, we further analyze the influence of edge supervision in different levels on segmentation performance by adding or reducing level edge supervision in the low-level features.  
To facilitate the analysis, ResUNet with Co-supervision and Fusion Model (ResUNet\_C$_i$F) represents the first $i$ levels (i.e., $S_1,\dots, S_i$) in the low level to use ESM, while the rest (i.e., $S_{i+1},\dots,S_n$) adopt ASSM in the Co-supervision, where $n$ is the number of down-sampling ($n=5$ here). The results is illustrated in Table $\ref{tab3}$, and it is obvious that Dice coefficient firstly rises and then declines along the first level number $i$ from $1$ to $5$. When $i=2$ (i.e., ResUNet\_C$_2$F), the proposed method can obtain the best segmentation performance. It means that the features of low-level boundary and high-level semantic can both be strengthened as the first level number $i$ increases and reduces, respectively. When $i=2$, there is a trade-off between the number of low-level and high-level (i.e., the use of context and localization accuracy), consequently ResUNet\_C$_2$F can surpass other ResUNet\_C$_i$F in most  metrics, such as $Dice$, $MAE$, $E_{\phi}$ and $S_{\alpha}$. More precisely, the proposed ESM and ASSM can incorporate low-level details with high-level semantics from feature maps in different levels by using AFM.
\subsubsection{Qualitative Results}
To further demonstrate the effectiveness of the proposed scheme, we visualize the prediction results of different networks. As shown in Figure $\ref{fig_6}$, our method can  remarkably outperform the baseline methods in the lung infection segmentation. Specifically, our segmentation results have much less mis-segmented tissues, while there are a lot of lossing and improper segmentation in the baseline U-Net and other methods. For the infection edge marked with a red box, for instance, our method can obtain a complete edge, and it is much closer to the real label in edge detail, which benefits from the more detailed edge information provided by the proposed ESM. Besides, from the regions marked by the blue box, our method can avoid over-segmentation, under-segmentation and incorrect segmentation efficiently. Especially in the $4^{th}$ rows, only our method and Deeplabv3+ can correctly detect the small infection (marked the blue box). It can be also observed obviously that our method is better than Deeplabv3+ in the edge details of large targets (marked the red box) because our method can provide different sizes of receptive fields and have good segmentation performance for different scale objects.\par

Along the down-sampling process in U-Net, edge feature information becomes less and less, while semantic one becomes richer and richer. For further verification, we visualize the feature maps of different levels (i.e., from $S_1$ to $S_5$) in  ResUNet\_C$_5$F. As shown in Figure $\ref{fig_7}$, the feature maps of low-level output ($S_1$ and $S_2$) contain more details, and the feature map in $S_3$ is the closest to the edge GT. With the deepening of down-sampling, edge feature information becomes less obvious. In the back propagation, we can extract more semantic information from the feature maps of high-level, as shown in $S_5$. It demonstrates that our ESM in low-level and ASSM in hige-level are very efficient to deal with a such difficult segmentation. 

\begin{table*}[]
\begin{center}
\small
\caption{The results of different numbers of edge supervised on our dataset. Bold black text and blue text represent the first and second best results, respectively.}
\begin{tabular}{lllllll}
\toprule[1pt]  
Methods& $Dice(\%)\uparrow$  &$Sens.(\%)\uparrow$  &$Prec.(\%)\uparrow$  & $MAE(\%)\downarrow$  & $E_{\phi}$(\%)$\uparrow$  &$S_{\alpha}$(\%)$\uparrow$    \\  \midrule[1pt] 
 ResUNet\_C$_1$F & 89.16$\pm$0.49   & 88.03$\pm$1.49     & \textbf{92.08$\pm$1.05} & 0.56$\pm$0.03  &95.27$\pm$0.12  &\textcolor{blue}{85.59$\pm$0.80}   \\
 ResUNet\_C$_2$F  & \textbf{89.93$\pm$0.09} &\textcolor{blue}{90.29$\pm$0.66} &\textcolor{blue}{91.91$\pm$0.97}& \textbf{0.52$\pm$0.01}  & \textbf{95.69$\pm$0.10} & \textbf{86.75$\pm$0.07}  \\ \hline
 ResUNet\_C$_3$F &\textcolor{blue}{89.44$\pm$0.14} & 90.15$\pm$0.88     & 91.90$\pm$1.13  &\textcolor{blue}{0.55$\pm$0.01}&95.30$\pm$0.49  & 85.41$\pm$0.09   \\
 ResUNet\_C$_4$F & 89.40$\pm$0.33   &\textbf{90.66$\pm$0.45}     & 91.12$\pm$0.92  & 0.58$\pm$0.02  &\textcolor{blue}{95.32$\pm$0.24}& 85.35$\pm$0.43   \\
 ResUNet\_C$_5$F & 88.33$\pm$0.89   & 90.28$\pm$0.67     & 90.05$\pm$0.45  & 0.58$\pm$0.06  &95.05$\pm$1.30  & 85.29$\pm$1.58   \\
\bottomrule[1pt]  
\end{tabular}
\label{tab3}
\end{center}
\end{table*}

\begin{figure*}[htbp]
\begin{center}
\includegraphics[width=0.9\textwidth]{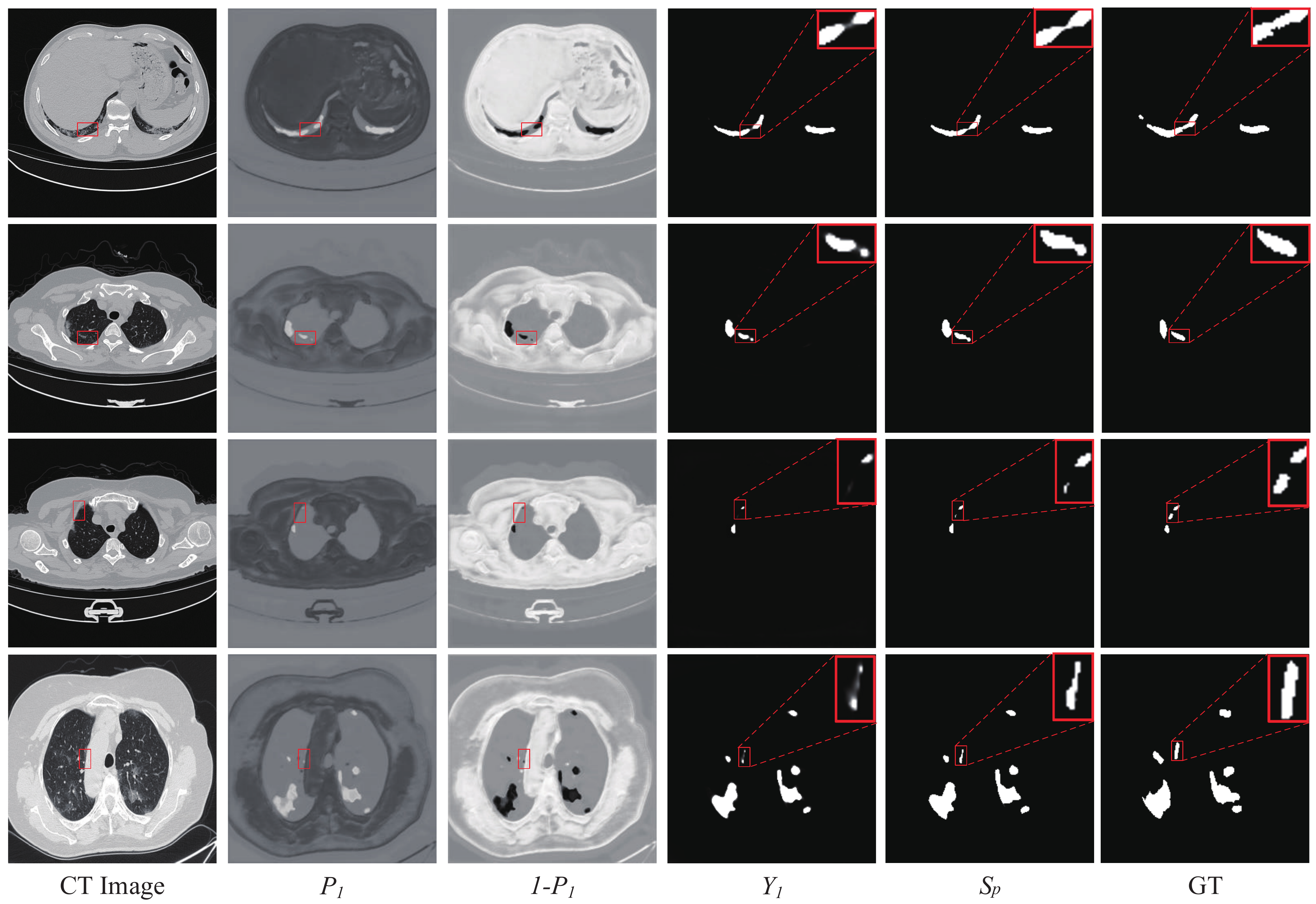}
\caption{Visual results of the fusion process based on the proposed AFM. Column 1: the original CT image; Column 2: the obtained confidence map $P_1$; Column 3: the confidence map $1-P_1$ of the lost detailed information; Column 4: the major result $Y_1$ from the top feature map $x_1$ through the attention block (AB); Column 5: the final prediction result $S_p$; Column 6: the corresponding ground truth (GT).}
\label{fig_8}
\end{center}
\end{figure*}
\subsection{Ablation Experiments}
To further analyse and test the validity of the proposed modules, a series of comparison experiments are conducted on our dataset by using various combinations among ESM, ASSM  and AFM based on the baseline ResUNet. The experimental results are shown in the third row of Table \ref{tab4}, and each module can improve independently the Dice coefficient of infection segmentation. Thereinto, compared with the baseline ResUNet without any other modules, ASSM can obtain independently the greatest performance improvements, followed by AFM. While for various combinations between ESM, ASSM and AFM, they can also outperform their separate modules, and the combination of ASSM and AFM can obtain slightly better performance than that of  ESM and AFM. Finally, the combination of the three modules can obtain the best performance, the reason is that the integration can take full advantage of them and obtain the optimal segmentation effect. Our network can be generalized for other segmentation applications due to the effectiveness of its architecture.\par

To test the effects of the proposed modules in the decoder, ESM and ASSM are applied separately or jointly in the up-sampling path. For convenience, $^*$ indicates the corresponding modules and stages in the up-sampling path (shown in Figure \ref{fig_2}). Owing the symmetric structure between the encoder and decoder, ESM and ASSM are symmetrically placed in the low level (i.e., $S^*_1$ to $S^*_2$) and high level (i.e., $S^*_3$ to $S^*_5$)  of the up-sampling path, respectively. The experimental results are shown in the fourth row of Table \ref{tab4}. Compared with the baseline method, the Dice performance can be improved in certain extent when these modules are separately or jointly adopted in up-sampling path, particularly the combination of the three modules can obtain the second best segmentation performance. However, the obtain performance in the up-sampling path is slightly worse than that of the corresponding down-sampling path in general. It means that the proposed Co-supervision scheme can both guide the network learning the features of edges and semantics in the down-sampling and up-sampling paths, but the effect would be more appreciable when the supervision modules is applied in the down-sampling path. The reason is that the levels of the down-sampling path contain richer primitive feature information than those of the up-sampling path owing to the encoder close to the original input data, while the edge and semantic information exist more or less some loss and noise when reconstructing a higher resolution layers by using bilinear interpolation up-sampling. Accordingly, the supervision in the levels of the down-sampling path is more stronger than that of the up-sampling path.\par


However, interestingly, the segmentation performance is even decreased compared with the baseline method when the proposed Co-supervision scheme are simultaneously applied in the down-sampling and up-sampling paths, and the fourth row of Table \ref{tab4} shows the results. Except the combination between ESM, ESM$^*$ and AFM, all combinations between the down-sampling and up-sampling paths can obtain poorer segmentation performance than the baseline method. While the combination between ESM, ESM$^*$ and AFM can increase by about 1\% over the baseline method. The most probable cause is the conflict and interference of the Co-supervision between the down-sampling and up-sampling paths. For example, the down-sampling path (i.e., encoder) is used to encode the input image into feature representations at multiple different levels, thereby capturing the context of the image like edge detail information. While the up-sampling path (i.e., decoder) is to semantically project the discriminative features (lower resolution) learned by the encoder onto the pixel space (higher resolution) to get a precise localization. Correspondingly, the loss function is to put more emphasis on edge details in the encoder path, while to highlight localization information for the decoder path. But all the feature maps of the decoder come from the encoder by concatenating and up-sampling, which results in the conflict and interference between the encoder and decoder when the Co-supervision modules are simultaneously applied in the two paths.
\begin{table*}[]
\begin{center}
\small
\caption{Ablation Experiments on our dataset. Bold black text and blue text represent the first and second best results, respectively.}
\begin{tabular}{l|cccccll|l}
\toprule[1pt]   
Baseline& ESM    &ASSM  &ASSM$^*$ & ESM$^*$  & AFM            &$Dice(\%)\uparrow$             \\ 
\midrule 
\multirow{17}{*}{ResUnet}
		&              &                &              &                   &                      & 85.96$\pm$0.03     \\
\cline{2-7}
         & $\surd$  &                &              &                   &                      & 87.08$\pm$0.45     \\
         &              & $\surd$    &              &                   &                      & 87.91$\pm$0.83         \\
         &              &                &              &                   & $\surd$          & 87.59$\pm$1.07         \\
         & $\surd$  &                &              &                   & $\surd$          & 88.33$\pm$0.89         \\ 
         &              & $\surd$    &              &                   & $\surd$          & 88.70$\pm$0.25              \\  
         & $\surd$  & $\surd$    &             &                    & $\surd$          & \textbf{89.93$\pm$0.09}  \\ 
\cline{2-7}  
         &              &                &              & $\surd$       &                      & 87.17$\pm$0.58     \\
         &              &                & $\surd$  &                   &                      & 86.47$\pm$0.46      \\
         &              &                & $\surd$  &                   & $\surd$          & 87.31$\pm$0.58      \\
         &              &                &              & $\surd$       & $\surd$          & 87.99$\pm$0.36      \\
         &              &                & $\surd$  & $\surd$       & $\surd$          &\textcolor{blue}{88.86$\pm$0.31}          \\  
\cline{2-7}   
         & $\surd$  &                &              & $\surd$       & $\surd$          & 86.95$\pm$0.37          \\   
         &              & $\surd$    & $\surd$  &                   & $\surd$          & 85.19$\pm$0.23          \\  
		& $\surd$  &                & $\surd$  &                   & $\surd$          & 85.55$\pm$0.47          \\ 
		&              & $\surd$    &              & $\surd$       & $\surd$          & 85.11$\pm$0.09          \\ 
         & $\surd$  & $\surd$    & $\surd$  & $\surd$       & $\surd$          & 85.63$\pm$0.51        \\        
\bottomrule[1pt] 
\end{tabular}
\label{tab4}
\end{center}
\end{table*}
\subsection{Comparison of Fusion Methods}
Multilevel feature fusion means different level of feature maps are integrated together to enrich the feature information, and traditional fusion approaches usually use feature addition or concatenation. An addition process is to add multiple feature maps to be one, which means that the amount of information under the characteristics of the description image is increased. While a concatenation is a combination of the number of channels, which means that the features describing the image itself are increased, but the information under each feature is not increased. To further verify the advantage of the proposed AFM, a series of comparison experiments are carried out by only using different fusion approaches, and the segmentation results are shown in Table \ref{tab5}. It can be seen that the proposed AFM can surpass the other two methods in all metrics except $S_{\alpha}(\%)$.  The reason is that all feature maps are evenly fused according to the same importance in the adding or concatenating process. However, it is obviously unreasonable because there are great differences between different levels in feature representations, and it is not sufficient to adaptively compensate low level finer details to high level semantic features only by simple adding or concatenating operation. Meanwhile, the concatenation operation can reduce the weight of the feature maps with poor semantics in the subsequent features in the convolution layer, while retaining rich semantic features in the channel. Whereas the addition operation can weaken the discrimination of features due to the simple pixel-wise summation for the feature maps. Therefore, the concatenation fusion method can surpass the addition operation.\par

Figure \ref{fig_8} illustrates the visual results of the fusion process by utilizing an attention mechanism. $Y_1$ is only processed by the attention block (AB), thereby it is the nearest output to the segmentation prediction of the baseline. While $S_p$ is the segmentation results by fusing multiple level feature maps, which would achieve the goal of both high resolution and rich semantics by combining the complementary strengths of multiple level feature maps. It is obvious that the $S_p$ is more complete than the $P_1$, and its lost information is  lesser than that of  the $P_1$. The obtain confidence $P_1$ attaches importance to the $P_1$ to ensure the most information retained. As a complement to the $P_1$, whereas, the confidence map $1-P_1$ pays attention to the lost detailed information, and it can exploit sufficient spatial and semantic features to supplement the lost detailed information  by fusing different levels. Thus the proposed methods can overcome the under-segmentation problem of the baseline, and retain multi-scale contextual information from multiple different levels.   \par

\begin{table*}[]
\begin{center}
\small
\caption{The results of different fusion methods on our dataset}
\begin{tabular}{lllllll}
\toprule[1pt]  
Methods& $Dice(\%)\uparrow$  &$Sens.(\%)\uparrow$  &$Prec.(\%)\uparrow$  & $MAE(\%)\downarrow$  & $E_{\phi}$(\%)$\uparrow$  &$S_{\alpha}$(\%)$\uparrow$    \\  \midrule[1pt] 
Add & 83.59$\pm$2.14   & 85.16$\pm$1.91     & 81.07$\pm$0.83  & 0.85$\pm$0.13  &93.66$\pm$0.10  &80.12$\pm$0.84   \\
Concatenate  & 86.75$\pm$1.38   & 87.00$\pm$0.92     & 86.93$\pm$1.03  & 0.64$\pm$0.08  & 94.39$\pm$1.04     & \textbf{84.12$\pm$1.27} \\ 
Attention & \textbf{87.59$\pm$1.07} & \textbf{88.04$\pm$1.12} & \textbf{87.18$\pm$1.36}& \textbf{0.59$\pm$0.05} & \textbf{95.05$\pm$1.30}& 83.89$\pm$1.21   \\\bottomrule[1pt] 
\end{tabular}
\label{tab5}
\end{center}
\end{table*}

\subsection{Comparisons on other COVID-19 Datasets}
To further verify the effectiveness and generalization ability, a series of comparison experiments are conducted on MosMedData [53], UESTC-COVID-19[52] and COVID-ChestCT [54], respectively. We select three important metrics for the evaluation of the COVID-19 lung infection segmentation, including $Dice$, $Sens.$ and $MAE$. The results are shown in Tables 6-8. For the MosMedData dataset, our method is slightly superior than Attention U-Net [34] and UNet++ [9] with Dice metric, but it can obtain 3.06\% better than its nearest competitor F3Net [20] with Sensitivity ($Sens.$), and can achieve the best performance among these three methods with all metrics (shown in Table 6). In the UESTC-COVID-19 dataset, our method is slightly better than its nearest competitor with $Dice$ and $MAE$ metrics, and is slightly lower than its nearest competitor in Sensitivity ($Sens.$). Overall, our method can obtain the best comprehensive performance among these methods (shown in Table 7). As for the COVID-ChestCT, our method can achieve the first, first and third best performance in Sensitivity ($Sens.$), $MAE$ and $Dice$, respectively. Compared with other methods, our method can also achieve the best overall performance (shown in Table 8). From the above results, our method can achieve the first three best performances for various datasets using all metrics, and has the best comprehensive performance comparing to other methods.
\begin{table*}[]
\begin{center}
\caption{Performance comparisons between different methods on MosMedData. Bold black text, blue text and green text represent the first, second and third best results, respectively.}
\small
\begin{tabular}{lccc}
\toprule[1pt]   
Methods         &$Dice(\%)\uparrow$        & $Sens.(\%)\uparrow$               &$MAE(\%)\downarrow$                \\ 
\midrule[1pt]   
U-Net \cite{7}                   & 80.39$\pm$9.865    &64.32$\pm$19.197     & 8.26$\pm$10.959        \\
UNet++ \cite{25}              &\textcolor{green}{87.27$\pm$2.102}     &74.80$\pm$7.508       &1.60$\pm$2.223            \\
Attention U-Net \cite{33}  &\textcolor{blue}{87.42$\pm$0.395}     &81.92$\pm$3.566      &0.30$\pm$0.059           \\
PSPNet \cite{11}              &82.38$\pm$1.048     &79.11$\pm$2.323       & 0.37$\pm$0.040          \\
Deeplabv3 \cite{12}          &83.70$\pm$1.166     &80.01$\pm$0.350      &0.34$\pm$0.026             \\
Inf-Net \cite{16}               &78.64$\pm$0.390     &78.17$\pm$4.277      &0.44$\pm$0.015           \\ 
SCRN \cite{28}                &87.13$\pm$0.070     &\textcolor{green}{82.42$\pm$3.957}    &\textcolor{blue}{0.26$\pm$0.008}            \\
F3Net \cite{Wei2020}       &83.66$\pm$0.337     &\textcolor{blue}{82.58$\pm$1.121}      &0.33$\pm$0.006           \\ 
DANet \cite{19}               &84.84$\pm$4.435     &70.65$\pm$8.948      &2.94$\pm$4.268            \\ 
ACFNet \cite{Zhang2019}&79.40$\pm$1.114     &79.07$\pm$3.978      &0.42$\pm$0.023          \\
CE-Net \cite{Gu2019}      &86.11$\pm$0.371     &76.17$\pm$0.121      & \textcolor{green}{0.28$\pm$0.012}           \\
CPFNet \cite{Feng2020}  &86.79$\pm$0.266     &73.50$\pm$4.910      &0.42$\pm$0.101         \\ \hline
ResUNet\_C$_2$F(\textbf{Ours}) &\textbf{87.43$\pm$0.165} &\textbf{85.64$\pm$3.013}  &\textbf{0.253$\pm$0.006}          \\ 
\bottomrule[1pt]    
\end{tabular}
\label{tab6}
\end{center}
\end{table*}
\begin{table*}[]
\begin{center}
\caption{Performance comparisons between different methods on UESTC-COVID-19. Bold black text, blue text and green text represent the first, second and third best results, respectively.}
\small
\begin{tabular}{lccc}
\toprule[1pt]   
Methods             & $Dice(\%)\uparrow$        & $Sens.(\%)\uparrow$                 & $MAE(\%)\downarrow$           \\ 
\midrule[1pt]   
U-Net \cite{7}                 &\textcolor{blue}{85.48$\pm$0.125}  &75.90$\pm$3.466         &0.47$\pm$0.060        \\
UNet++ \cite{25}            &85.13$\pm$0.296     &76.41$\pm$1.636         &\textcolor{blue}{0.47$\pm$0.010}         \\
Attention U-Net \cite{33}&84.62$\pm$0.539     &77.04$\pm$2.484         &0.50$\pm$0.035         \\
PSPNet \cite{11}            &82.67$\pm$0.248      &76.45$\pm$1.246        & 0.56$\pm$0.006        \\
Deeplabv3 \cite{12}       &80.13$\pm$1.212      &70.88$\pm$2.573        &0.63$\pm$0.049          \\
Inf-Net \cite{16}            &83.26$\pm$0.440      &77.45$\pm$1.810        &0.54$\pm$0.015        \\ 
SCRN \cite{28}              &83.78$\pm$0.402     &77.75$\pm$2.949        &0.52$\pm$0.017       \\
F3Net \cite{Wei2020}     &83.58$\pm$0.974     &78.56$\pm$1.178        &0.53$\pm$0.026        \\ 
DANet \cite{19}             &85.40$\pm$0.745     &\textcolor{green}{79.40$\pm$2.175}     &0.47$\pm$0.026         \\ 
ACFNet \cite{Zhang2019}&84.31$\pm$0.435   &\textbf{79.83$\pm$1.799}        &0.50$\pm$0.020          \\ 
CE-Net \cite{Gu2019}    &\textcolor{green}{85.45$\pm$0.420}       &77.38$\pm$7.962        &0.47$\pm$0.016       \\
CPFNet \cite{Feng2020} &85.36$\pm$0.182    &77.35$\pm$1.151         &\textcolor{green}{0.47$\pm$0.015}        \\ \hline
ResUNet\_C$_2$F(\textbf{Ours})&\textbf{85.52$\pm$0.081}    &\textcolor{blue}{79.46$\pm$2.286}   & \textbf{0.47$\pm$0.006}          \\
\bottomrule[1pt]    
\end{tabular}
\label{tab7}
\end{center}
\end{table*}
\begin{table*}[]
\begin{center}
\caption{Performance comparisons between different methods on COVID-ChestCT. Bold black text, blue text and green text represent the first, second and third best results, respectively.}
\small
\begin{tabular}{lccc}
\toprule[1pt]   
Methods                          &$Dice(\%)\uparrow$                 & $Sens.(\%)\uparrow$                & $MAE(\%)\downarrow$              \\ 
\midrule[1pt]   
U-Net \cite{7}                 & 71.86$\pm$0.240    &79.31$\pm$0.950      &\textcolor{blue}{0.75$\pm$0.012}      \\
UNet++ \cite{25}            &71.62$\pm$0.412     &72.77$\pm$5.927      &0.93$\pm$0.173        \\
Attention U-Net \cite{33}&70.44$\pm$1.420     &75.27$\pm$7.900      &0.95$\pm$0.303       \\
PSPNet \cite{11}            &65.95$\pm$1.979     &80.95$\pm$4.126      & 0.93$\pm$0.068    \\
Deeplabv3 \cite{12}        &59.41$\pm$1.213     &74.27$\pm$7.419      &1.05$\pm$0.095        \\
Inf-Net \cite{16}             &59.63$\pm$2.735     &78.32$\pm$0.131      &1.09$\pm$0.107      \\ 
SCRN \cite{28}              &68.46$\pm$1.345     &74.33$\pm$6.288      &1.15$\pm$0.471      \\
F3Net \cite{Wei2020}     &68.62$\pm$0.477     &\textcolor{blue}{83.30$\pm$5.989}      &0.83$\pm$0.031     \\ 
DANet \cite{19}             &71.10$\pm$0.731     &72.45$\pm$6.953      &0.92$\pm$0.182        \\ 
ACFNet \cite{Zhang2019}&68.69$\pm$1.979   &76.36$\pm$8.352      &0.84$\pm$0.042        \\ 
CE-Net \cite{Gu2019}    &\textcolor{blue}{73.72$\pm$0.583}     &71.38$\pm$1.666      &\textcolor{blue}{0.75$\pm$0.012}      \\
CPFNet \cite{Feng2020}&\textbf{74.65$\pm$1.163}     &\textcolor{green}{81.24$\pm$1.700}      &1.01$\pm$0.411      \\ \hline
ResUNet\_C$_2$F(\textbf{Ours})&\textcolor{green}{72.81$\pm$0.148}     &\textbf{83.89$\pm$1.358}      &\textbf{0.74$\pm$0.017}         \\ 
\bottomrule[1pt]    
\end{tabular}
\label{tab8}
\end{center}
\end{table*}
\section{Conclusion}
It is still a challenging task to accurately segment the infected lesions of COVID-19 on CT images owing to the irregular shapes with various sizes and indistinguishable boundaries between normal and infected tissues. In this paper, a novel segmentation scheme is proposed for the infection segmentation of COVID-19 on CT Images. To achieve this, we propose three modules for deep collaborative supervision and attention fusion based on ResUnet.  To verify the effectiveness of the proposed scheme, a series of experiments are conducted on four COVID-19 datasets. The results show that our method can achieve the best performance for most of the datasets with metrics, such as $Dice$, Sensitivity($Sens.$) and $MAE$, and has better generalization performance comparing to the existing approaches. \par

The proposed technique has four advantages as follows. Firstly, it is able to capture rich spatial information in various scales through  an edge supervised module, denoted as the ESM, which allows to incorporate the edge supervised information into the initial stage of down-sampling in the framework of ResUnet. As low-level layers contain richer object boundaries, they are used to define the edge supervised loss function to capture all spatial information. The main benefit of this module is to highlight low-level boundary features and provide useful fine-grained constraints to guide feature extraction in semantic segmentation tasks. Secondly, the proposed method can explore semantic information from various scale infections on COVID-19 CT images by using an auxiliary semantic supervised module (i.e., ASSM) that can integrate the appearance supervised information into the later stage of down-sampling. The main advantage of this module is to strengthen high-level semantic information during the feature extraction process. Thirdly, we propose an attention fusion module (i.e., AFM) to fuse multiple scale feature maps of different levels from the up-sampling stage to reduce the semantic gaps between high-level and low-level feature maps. The main advantage of this module is to strengthen and supplement the lost detailed information in high-level representations. Lastly, we construct a joint loss function by combining the edge supervised loss, auxiliary semantic supervised loss and fusion loss. The joint function can guide the network in learning the features of COVID-19 infections, thereby achieving a deep collaborative supervision on edges and semantics. Meanwhile, it can also act as an incentive to effectively fuse multi-scale feature maps of different levels.
\par

Although our network can get a good result in segmenting the overall infection region, it is not sufficient to estimate the severity of infected COVID-19, because finer segmentation of the different infection regions is required. In the future, we might collect a large amount of  COVID-19 data, and consider further recognizing the severity of COVID-19 according to the area, size, and location of infections. 

\section*{Acknowledgement}
The authors would like to express their appreciation to the referees for their helpful comments and suggestions.
This work was supported in part by Zhejiang Provincial Natural Science Foundation of China (Grant No. LGF20H180002), and in part by National Natural Science Foundation of China (Grant No. 61802347, 61801428 and 61972354), the National Key Research and Development Program of China (Grant No. 2018YFB1305202), and the Microsystems Technology Key Laboratory Foundation of China.

\bibliographystyle{unsrtnat}

\bibliographystyle{elsarticle-num}





\end{document}